# Research on Portfolio Liquidation Strategy under Discrete Times

Qixuan Luo[1], Yu Shi[1], Handong Li[1]

(1. School of Systems Science, Beijing Normal University, Beijing, China)

**Abstract**

This paper presents an optimal strategy for portfolio liquidation under discrete time conditions. We assume that N risky assets held will be liquidated according to the same time interval and order quantity, and the basic price processes of assets are generated by an N-dimensional independent standard Brownian motion. The permanent impact generated by an asset in the portfolio during the liquidation will affect all assets, and the temporary impact generated by one asset will only affect itself. On this basis, we establish a liquidation cost model based on the VaR measurement and obtain an optimal liquidation time under discrete-time conditions. The optimal solution shows that the liquidation time is only related to the temporary impact rather than the permanent impact. In the simulation analysis, we give the relationship between volatility parameters, temporary price impact and the optimal liquidation strategy.

## 1. Introduction

Trade execution strategies adopted by financial institutions and hedge funds have attracted more and more attention from academia. A key issue for institutional traders is how to settle large stock orders properly. In the case of quick trading, traders have to bear the risk of more costs because of market impact, which reflects the market liquidity. Thus, to minimize price impact, investors may divide large orders into small pieces. However, the longer liquidation time will lead to higher risks owing to price volatility during the investment horizon. Hence, an excellent execution strategy is to adjust the transactions appropriately to achieve a tradeoff of reducing transaction costs from price impact and lowering volatility risk taken. These problems have been studied by Bertsimas and Lo [1], Almgren and Chriss [2], Cetin, Jarrow and Protter [3], Schied and Schöneborn [4], Kharroubi and Pham [5], and Obizhaeva and Wang [6].

In the real world, institutional investors usually have to think over the problems of portfolio liquidation. Large transactions consume liquidity in the market, resulting in higher execution costs. Therefore, it is very important to arrange order flows in an appropriate way to balance the gains and costs of liquidation. In the research by Almgren and Chriss [2], minimizing the combination of trading costs caused by market impact and volatility risk was considered.

Gatheral and Schied [7] indicated that the price impact models can be generally divided into two categories. The first one included temporary impact and permanent impact, which the former only affected current transaction and the latter would have an impact on the current and all future trades. The second one emphasized the transience of price impact, which arose from large transactions but would gradually subside once they were executed. These models aimed at splitting the individual orders in an optimal manner. Recently, the problems of dynamic portfolio related to price impact are becoming more and more concerned, which means the orders to be liquidated are derived from a dynamic optimization problem endogenously and are no longer supposed to be given. This can be modeled by considering the balance of gains from reaction to new information and transaction costs. However, the complexity of the optimization problems is greatly increased. Hence, people's attention is almost completely concentrated on the first-generation price impact models. So far, the transaction costs of these models are only temporary.

However, due to the interrelationship between assets, the process of solving the liquidation of multiple risky assets is very complicated, and its value in practical application is greatly reduced. From the perspective of practical application, this paper discusses the optimal liquidation time of consecutive orders according to fixed time intervals and equal order quantity under a given portfolio.

The remainder of this study is organized into the following sections. In the next section we propose the theoretical framework, and establish the impact cost model under certain assumptions and give the optimal liquidation time. The third section reviews the simulation analysis, which discusses the effects of initial position, volatility and temporary impact on the liquidation strategy of the portfolio. The last section concludes.

## 2. The liquidation model and the optimal solution

Assuming that an investor holds a portfolio of N risky assets, and the initial asset position is $x_0 = (x_0^1, x_0^2, \cdots, x_0^N)' \in \mathbb{R}^N$, which needs to be liquidated (sold) at the same time within time T. Suppose that the portfolio position held at t($0 \leq t \leq T$) is $x_t = (x_t^1, x_t^2, \cdots, x_t^N)' \in \mathbb{R}^N$. Given the boundary condition $x_T = 0$ and assuming that $x_t$ is continuous with respect to t, for the $ith(i = 1,2,\cdots,N)$ asset, there is $\frac{dx_t^i}{dt} = -v_t^i$, where $v_t^i$ is the liquidation speed of the ith asset. Hence

$$x_t = x_0 - \int_0^t v_s ds \tag{1}$$

Due to market liquidity constraints and other factors, investors' trading behavior will cause changes in asset prices. We define the price vector of N assets as $\widetilde{S}_t \in \mathbb{R}^N$, and the price can be divided into two parts, namely, the basic part of price $S_t$ and the impact part $Im_t$, thus,

$$\widetilde{S}_t = S_t - Im_t \tag{2}$$

According to the analysis of price impact by Hlothausen, Leftwich and Mayers [8], we divide price impact into permanent market impact and temporary market impact. It is assumed that there is a linear relationship between the permanent market impact and the number of shares sold and that the permanent impact among assets will affect all assets during the entire period of transactions. At the same time, we assume that there is a linear relationship between the temporary market impact and the speed of selling stocks, which only affects the current transaction and one asset. It can be written as

$$Im_t = \int_0^t PermIm(v_s)\, ds + TempIm(v_t) \in \mathbb{R}^N$$

where

$$PermIm(v_s) = \gamma v_s$$

$$TempIm(v_t) = \eta v_t$$

where $PermIm(\cdot)$ represents permanent impact, and $TempIm(\cdot)$ represents temporary impact, and $\gamma = (\gamma^{ij}) \in \mathbb{R}^{N \times N}$ represents the permanent impact coefficient matrix, and $\eta = diag(\eta^1, \eta^2, \cdots, \eta^N)$ represents the temporary impact coefficient matrix.

For the convenience of analysis, it is generally assumed that the N-dimensional basic price process $S_t$ is subject to the N-dimensional random walk model. Therefore, the asset price dynamics equation is defined as:

$$\widetilde{S}_t = S_0 + \sigma B_t - \gamma \int_0^t v(s)\, ds - \eta v_t$$

which is

$$\widetilde{S}_t = S_0 + \sigma B_t - \gamma(x_0 - x_t) - \eta v_t \tag{3}$$

where $S_0 \in \mathbb{R}^N$ denotes the asset initial price vector and $B_t$ denotes the N-dimensional standard Brownian motion. $\sigma = (\sigma^{ij})_{N \times N}$, where $\sigma^{ij}, i,j = 1,2,\cdots,N$ is the volatility coefficient of the jth Brownian motion component to the ith asset price.

Then, the execution price process of the ith asset is:

$$\widetilde{S}_t^i = S_0^i + \sigma^{i\cdot} B_t - \gamma^{i\cdot}(x_0 - x_t) - \eta^i v_t^i \tag{4}$$

where $\sigma^{i\cdot} = (\sigma^{ij})_{1 \times N}$, $\gamma^{i\cdot} = (\gamma^{ij})_{1 \times N}, i = 1, \cdots, N$.

It can be seen from equation (4) that the process of each asset price consists of two parts, which are the basic price process and the price impact caused by transactions. The basic price process is generated by a common random source—the N-dimensional standard Brownian motion. The price impact also includes two aspects, that is, the price process of an asset is not only affected by the permanent price impact caused by its own transactions and other assets' transactions but also by the temporary price impact caused by the transactions of the asset itself.

For the given volatility matrix $\sigma$, if the matrix is a lower triangular matrix and satisfies $\Sigma = \sigma \sigma^T \in \mathbb{R}^{N \times N}$, where $\Sigma$ is positive definite, the $\Sigma$ is the covariance matrix of N assets, and the $\sigma$ is called the Cholesky decomposition of the N-dimensional asset covariance matrix.

Next, we discuss the optimal liquidation under the discrete time frame.

Under the framework of discrete time, the time interval is given by $\tau$ and $M = [T/\tau]$, where $[\cdot]$ denotes the integer part. The residual number of shares held in each time is defined as $x_0, x_1, \cdots, x_M$, and the residual amount of assets positions at time k is $x_k = \left(x_k^1, x_k^2, \cdots, x_k^N\right)'$. Thus, the number of shares sold per unit time in the kth period is:

$$v_k = \frac{\delta_k}{\tau}$$

where $\delta_k = x_{k-1} - x_k$, k = 1, 2 $\cdots$, M. According to equation (3), there is

$$\widetilde{S_k} = S_{k-1} + \tau^{\frac{1}{2}}\sigma\xi_k - \gamma(x_{k-1} - x_k) - \eta v_k = S_0 + \tau^{\frac{1}{2}}\sigma\sum_{j=1}^{k}\xi_j - \gamma\sum_{j=1}^{k}(x_{j-1} - x_j) - \eta v_k \quad (5)$$

where $\xi_k$ is a random variable following the standard normal distribution. At this point, we can obtain the expression for the total liquidation cost C.

**Theorem 1.** The total cost C of N assets liquidation can be expressed as:

$$C = -\tau^{\frac{1}{2}}\sum_{k=1}^{M} x_{k-1}'\sigma\xi_k + \frac{1}{2}x_0'\gamma x_0 + \sum_{k=1}^{M}(x_{k-1} - x_k)'\left(\frac{1}{2}\gamma + \frac{\eta}{\tau}\right)(x_{k-1} - x_k) \quad (6)$$

*Proof:* From the initial conditions and equation (5), there is:

$$C = x_0'S_0 - \sum_{k=1}^{M}(x_{k-1} - x_k)'\widetilde{S_k}$$

$$= x_0'S_0 - \left[\sum_{k=1}^{M}(x_{k-1} - x_k)'S_0 + \sum_{k=1}^{M}(x_{k-1} - x_k)'\tau^{\frac{1}{2}}\sigma\sum_{j=1}^{k}\xi_j - \sum_{k=1}^{M}(x_{k-1} - x_k)'\gamma(x_0 - x_k)\right.$$

$$\left. - \sum_{k=1}^{M}(x_{k-1} - x_k)'\cdot\frac{\eta}{\tau}\cdot(x_{k-1} - x_k)\right]$$

$$= -\tau^{\frac{1}{2}}\sum_{k=1}^{M} x_{k-1}'\sigma\xi_k + \frac{1}{2}x_0'\gamma x_0 + \frac{1}{2}\sum_{k=1}^{M}(x_{k-1} - x_k)'\cdot\gamma\cdot(x_{k-1} - x_k)$$

$$+ \sum_{k=1}^{M}(x_{k-1} - x_k)'\cdot\frac{\eta}{\tau}\cdot(x_{k-1} - x_k)$$

$$= -\tau^{\frac{1}{2}}\sum_{k=1}^{M} x_{k-1}'\sigma\xi_k + \frac{1}{2}x_0'\gamma x_0 + \sum_{k=1}^{M}(x_{k-1} - x_k)'\left(\frac{1}{2}\gamma + \frac{\eta}{\tau}\right)(x_{k-1} - x_k)$$

□

The total liquidation cost C is a random variable, which we can further obtain its expectation and variance. Hence, we have the following theorems.

**Theorem 2.** If the cost of portfolio liquidation is given by equation (6), the expectation and variance of cost are, respectively,

$$E[C] = \frac{1}{2}x_0'\gamma x_0 + \sum_{k=1}^{M} (x_{k-1} - x_k)' \left(\frac{1}{2}\gamma + \frac{\eta}{\tau}\right)(x_{k-1}-x_k) \qquad (7)$$

$$V[C] = \tau \sum_{k=1}^{M} x_{k-1}' \sigma \sigma' x_{k-1} \qquad (8)$$

*Proof:* Since the cost C is a function of the random variable $\xi$, it is easy to derive equation (7). Now we prove equation (8).

$$V[C] = E[C - E(C)]^2 = E\left[-\tau^{\frac{1}{2}} \sum_{k=1}^{M} x_{k-1}' \sigma \xi_k\right]^2 = \tau \sum_{k=1}^{M} E[x_{k-1}' \sigma \xi_k][x_{k-1}' \sigma \xi_k]'$$

$$= \tau \sum_{k=1}^{M} x_{k-1}' \sigma E[\xi_k \cdot \xi_k'] \sigma' x_{k-1} = \tau \sum_{k=1}^{M} x_{k-1}' \sigma \cdot I \cdot \sigma' x_{k-1} = \tau \sum_{k=1}^{M} x_{k-1}' \sigma \sigma' x_{k-1}$$

□

Given a liquidation strategy $\{\tilde{x}\} = (\widetilde{x_0}, \widetilde{x_1}, \cdots, \widetilde{x_M})$, where $\widetilde{x_k}$ is the vector of positions remaining of N assets at time $k (k = 1, \cdots, M)$. Since the orders placed in the liquidation process are equally spaced, the solution to the optimal liquidation strategy can also be translated into the solution to the optimal time length T, that is, $T = \tau M$, where $\tau$ is the time interval for placing orders.

We assume that the optimal execution strategy of the portfolio depends on minimizing the cost of liquidating the portfolio's position. The cost is considered in the sum of the expected execution cost and the cost of the market risk assumed.

We define $\text{VaR}_p(M)$ as the value at risk at the p confidence level, namely, the transaction cost the strategy may cause during the realization period. Then, an optimization model can be established as follows,

$$\text{VaR}_p(M) = E[C] + Z_p\sqrt{V[C]}$$

$$= \frac{1}{2}x_0'\gamma x_0 + \sum_{k=1}^{M}(x_{k-1} - x_k)'\left(\frac{1}{2}\gamma + \frac{\eta}{\tau}\right)(x_{k-1}-x_k) + Z_p\sqrt{\tau \sum_{k=1}^{M} x_{k-1}'\sigma\sigma' x_{k-1}} \qquad (9)$$

where $Z_p$ is the p quantile of the standard normal distribution, indicating that the execution of orders according to this strategy can make the transaction cost of assets in the process of realization not exceed $\text{VaR}_p(M)$ with the probability of p. Hence, the optimal liquidation strategy can be converted into the strategy when the solution of (9) reaches the minimum value.

According to the research results of Bertsimas and Lo [1], under the assumption that the market impact function is a linear function and the asset price follows the arithmetic random walk process, the strategy of liquidating stocks at a constant rate is the optimal strategy.

**Theorem 3.** Optimization: the optimal execution time based on $\text{VaR}_p(M) = E[C] + Z_p\sqrt{V[C]}$ is:

$$T^* = \left(\frac{2\sqrt{3}x_0'\eta x_0}{Z_p\sqrt{x_0'\sigma\sigma'x_0}}\right)^{\frac{2}{3}} \qquad (10)$$

*Proof:* Given the time interval $\tau$, the implicit expression of M can be obtained by (7) and (8), that is,

$$VaR_p(M) = \frac{1}{2}x_0'\gamma x_0 + \frac{x_0'\gamma x_0}{2M} + \frac{x_0'\eta x_0}{\tau M} + Z_p\sqrt{\frac{\tau x_0'\sigma\sigma'x_0 M}{3}\left(1+\frac{1}{M}\right)\left(1+\frac{1}{2M}\right)}$$

Taking the derivative of this and setting the derivative to zero, we can obtain:

$$\frac{\partial VaR_p(M)}{\partial M} = -\frac{x_0'\gamma x_0}{2M^2} - \frac{x_0'\eta x_0}{\tau M^2} + Z_p\sqrt{\frac{\tau x_0'\sigma\sigma'x_0 M}{3}}\cdot\frac{1-\frac{1}{2M^2}}{2\sqrt{M+\frac{1}{2M}+\frac{3}{2}}} = 0$$

that is,

$$\frac{\sqrt{3}x_0'\left(\frac{\gamma}{2}+\frac{\eta}{\tau}\right)x_0}{Z_p\sqrt{\tau x_0'\sigma\sigma'x_0}} = \frac{M^2-\frac{1}{2}}{2\sqrt{M+\frac{1}{2M}+\frac{3}{2}}}$$

Since $\tau = T/M$, the above equation can be denoted as follows,

$$\frac{\sqrt{3M}x_0'\left(\frac{\gamma}{2}+\frac{\eta M}{T}\right)x_0}{Z_p\sqrt{Tx_0'\sigma\sigma'x_0}} = \frac{M^2-\frac{1}{2}}{2\sqrt{M+\frac{1}{2M}+\frac{3}{2}}}$$

Then, we take the limit of the discrete time so that $\tau \to 0$, $M \to \infty$, and make the model frame approach from the discrete time to the continuous time. We can get the expression of the optimal execution time under this condition as

$$T^* = \left[\frac{2\sqrt{3}x_0'\eta x_0}{Z_p\sqrt{x_0'\sigma\sigma'x_0}}\right]^{\frac{2}{3}}$$

□

Theorem 3 shows that the optimal liquidation strategy of a portfolio is to determine the total time T of order liquidation under the condition of placing orders at equal intervals and equal quantities. The result shows that the optimal strategy $T^*$ for portfolio liquidation is only related to the initial position, the temporary impact coefficients and the volatility coefficients of assets, and has nothing to do with the permanent impact of each asset.

3. **Simulation analysis of liquidation strategy**

To discuss the impact of different parameters of the model on the optimal liquidation strategy, we simulate and analyze the liquidation problem of the portfolio composed of two assets. Based on the results obtained above, we suppose the price processes of the two assets are respectively,

$$\widetilde{S_k^1} = S_{k-1}^1 + \tau^{\frac{1}{2}}\sigma^{11}\xi_k^1 + \tau^{\frac{1}{2}}\sigma^{12}\xi_k^2 - \gamma^{11}(x_{k-1}^1 - x_k^1) - \gamma^{12}(x_{k-1}^2 - x_k^2) - \eta^1 v_k^1$$

$$\widetilde{S_k^2} = S_{k-1}^2 + \tau^{\frac{1}{2}}\sigma^{21}\xi_k^2 + \tau^{\frac{1}{2}}\sigma^{22}\xi_k^1 - \gamma^{22}(x_{k-1}^2 - x_k^2) - \gamma^{21}(x_{k-1}^1 - x_k^1) - \eta^2 v_k^2$$

where $\sigma^{11}, \sigma^{12}$ and $\sigma^{21}, \sigma^{22}$ are the two assets' volatility components, respectively. Therefore, the volatilities of the two assets are:

$$\sigma^1 = \sqrt{(\sigma^{11})^2 + (\sigma^{12})^2}$$

$$\sigma^2 = \sqrt{(\sigma^{21})^2 + (\sigma^{22})^2}$$

Regardless of the impact of asset prices, the correlation between the two asset price processes can be expressed as a correlation coefficient:

$$\rho = \frac{\sigma^{11}\sigma^{22} + \sigma^{12}\sigma^{21}}{\sigma^1 \cdot \sigma^2}$$

According to theorem 3, the optimal liquidation time is

$$T^* = \left[\frac{2\sqrt{3}\left([(x_0^1)^2\eta^1 + (x_0^2)^2\eta^2]\right)}{Z_p\sqrt{(x_0^1)^2[(\sigma^{11})^2 + (\sigma^{12})^2] + 2x_0^1 x_0^2(\sigma^{11}\sigma^{21} + \sigma^{12}\sigma^{22}) + (x_0^2)^2[(\sigma^{21})^2 + (\sigma^{22})^2]}}\right]^{\frac{2}{3}}$$

And the corresponding liquidation costs of the two assets are respectively:

$$C^1 = x_0^1 S_0^1 - \sum_{k=1}^{M}(x_{k-1}^1 - x_k^1)\widetilde{S_k^1}$$

$$C^2 = x_0^2 S_0^2 - \sum_{k=1}^{M}(x_{k-1}^2 - x_k^2)\widetilde{S_k^2}$$

Meanwhile, for the convenience of analysis, we define the total liquidation cost rate as:

$$CP^w = \frac{C^1 + C^2}{x_0^1 S_0^1 + x_0^2 S_0^2}$$

We suppose the positions of two assets that an investor needs to liquidate within time T are 10 million shares and 8 million shares, respectively, and the initial prices are 50 dollars and 100 dollars, respectively. We use the simulation method to analyze the relationship between temporary impact coefficients, volatility coefficients, optimal liquidation time, and liquidation costs in the portfolio liquidation strategy, and discuss the relationship between the permanent impact coefficients and the liquidation costs. During the simulation, we repeat the experiment for 1,000 times in each case and count the results. We take the mean value of all the total liquidation cost rate $CP^w$s obtained by repeated experiments in each case and get the average total liquidation cost rate $MCP^w$.

### 3.1 The effect of the temporary impact coefficients on asset liquidation

We first simulate and analyze the effect of the temporary impact coefficients on optimal liquidation time. The settings of temporary impact coefficients are shown in Table 1.

Table 1 Settings of temporary impact coefficients

|  | Initial Value | Adjustment Range | Adjustment Times | Final Value |
| --- | --- | --- | --- | --- |
| $\eta^1$ | $3\times10^{-8}$ | $3\times10^{-9}$ | 10 | $6\times10^{-8}$ |
| $\eta^2$ | $5\times10^{-8}$ | $5\times10^{-9}$ | 10 | $1\times10^{-7}$ |

We set the initial values of the temporary impact coefficients $\eta^1, \eta^2$ to $3\times10^{-8}$ and $5\times10^{-8}$, respectively, and gradually increase the values by 10% of the initial values and adjust them 10 times, thus obtaining 121 cases. Other parameters remain the same during the simulation. The volatility parameters $\sigma^{11}, \sigma^{12}$ of asset 1 are 0.08 and 0.02, respectively, and the volatility parameters $\sigma^{21}, \sigma^{22}$ of asset 2 are 0.1 and 0.03. The permanent impact coefficients $\gamma^{11}, \gamma^{22}$ of the two assets themselves are $3\times10^{-9}$ and $5\times10^{-9}$, respectively, and the external permanent impact coefficients $\gamma^{12}, \gamma^{21}$ are $1\times10^{-9}$ and $2\times10^{-9}$. In each case we run 1,000 simulations and count the results.

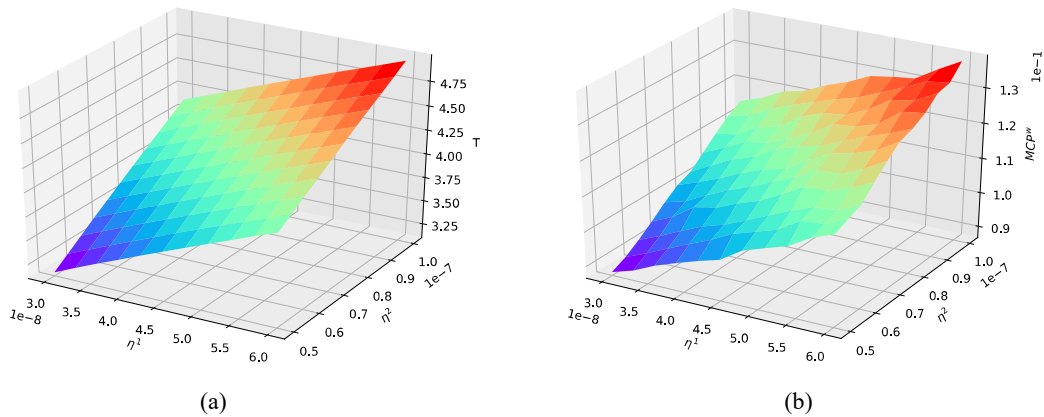

(a)　　　　　　　　　　　　　　(b)

Fig. 1 Relationship between the temporary impact coefficients and optimal liquidation time and average total liquidation cost rate.

In Fig. 1, the x-axis and y-axis represent the changes in the temporary impact parameters of the two assets, respectively, while the z-axis in Fig. 1(a) represents the optimal liquidation time and the z-axis in Fig. 1(b) represents the average total liquidation cost rate.

Fig. 1 shows that the optimal liquidation time and average total liquidation cost rate increase with the increase of the temporary impact coefficients of the two assets, showing a significant positive correlation. When $\eta^1, \eta^2$ are $3\times10^{-8}$ and $5\times10^{-8}$, respectively, the optimal liquidation time and average total cost rate reach the minimum of 3.14 and 8.77%. When $\eta^1, \eta^2$ are $6\times10^{-8}$ and $1\times10^{-7}$, respectively, the optimal liquidation time and average total cost rate reach the maximum of 4.98 and 13.79%.

This reflects that with the increasing temporary impact on the market, the optimal strategy will tend to be a longer time, and the cost of liquidation will gradually increase.

**3.2 The effect of volatility parameters on asset liquidation**

Next, we discuss the effect of volatility parameters on optimal liquidation time and cost.

**3.2.1 The effect of volatility components of single asset on asset liquidation**

First, we discuss the impact of volatility parameters on a single asset.

Table 2 Settings of volatility components of asset 1

|  | Initial Value | Adjustment Range | Adjustment Times | Final Value |
|---|---|---|---|---|
| $\sigma^{11}$ | 0.08 | 0.008 | 10 | 0.16 |
| $\sigma^{12}$ | 0.02 | 0.002 | 10 | 0.04 |

The settings of volatility components of asset 1 are shown in table 2. We set the initial values of the volatility components $\sigma^{11}, \sigma^{12}$ to be 0.08 and 0.02, respectively, and gradually increase the values by 10% of the initial values and adjust them 10 times, thus obtaining 121 cases. Other parameters remain unchanged, namely, the temporary impact coefficients $\eta^1, \eta^2$ are $3\times10^{-8}$ and $5\times10^{-8}$, the volatility components $\sigma^{21}, \sigma^{22}$ of asset 2 are 0.1 and 0.03, the permanent impact coefficients of themselves $\gamma^{11}, \gamma^{22}$ are $3\times10^{-9}$ and $5\times10^{-9}$, and the external permanent impact coefficients $\gamma^{12}, \gamma^{21}$ are $1\times10^{-9}$ and $2\times10^{-9}$. Each case is simulated 1,000 times and the results are shown in Fig. 2.

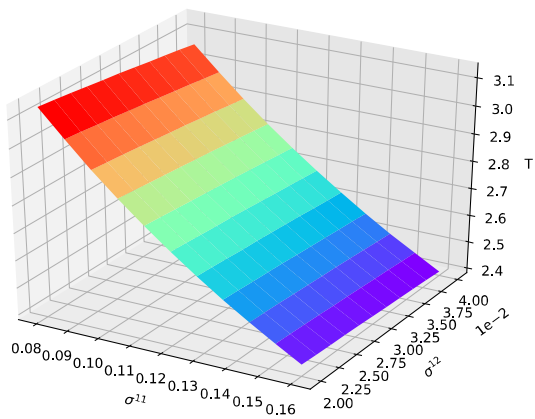 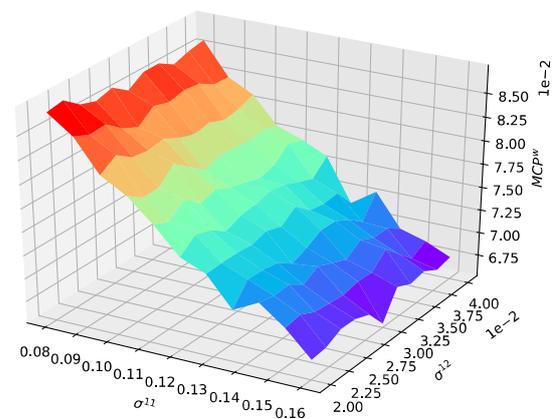

Fig. 2(a) Relationship between volatility components of asset 1 and optimal liquidation time

Fig. 2(b) Relationship between volatility components of asset 1 and average total liquidation cost rate

Fig. 2 Relationship between the volatility components of asset 1 and the optimal liquidation time and average total

liquidation cost rate

In Fig. 2, the x-axis and the y-axis represent $\sigma^{11}, \sigma^{12}$, respectively, while the z-axis in Fig. 2(a) represents the optimal liquidation time and the z-axis in Fig. 2(b) represents the average total liquidation cost rate.

Fig. 2 shows that the optimal liquidation time and average total liquidation cost rate generally decrease with the increase of the volatility components of asset 1, showing a significant negative correlation, which the influence of the component $\sigma^{11}$ is greater than the effect of the component $\sigma^{12}$. When $\sigma^{11}, \sigma^{12}$ are 0.16 and 0.04, the optimal liquidation time reaches the minimum of 2.397, and when $\sigma^{11}, \sigma^{12}$ are 0.08 and 0.02, the optimal liquidation time reaches the maximum of 3.1365. When $\sigma^{11}, \sigma^{12}$ are 0.16 and 0.03, the average total cost rate reaches the minimum of 6.56%, and when $\sigma^{11}, \sigma^{12}$ are 0.08 and 0.022, the average total cost rate reaches the maximum of 8.75%.

The above results show that as the market volatility increases, the optimal strategy tends to shorten the execution time, and the liquidation cost decreases gradually.

Simulation analysis of volatility components of asset 2 shows the same results, which are not listed here.

**3.2.2 The effect of correlation coefficient of two assets on asset liquidation**

In the process of portfolio liquidation, the relationship between asset prices is the core issue. On the basis of the previous analysis, the factors that affect the portfolio liquidation include the correlation of asset price volatility and the interaction of permanent price impact. The permanent price impact is deterministic and does not affect the final liquidation strategy. Hence, we focus on analyzing the impact of the relevance of the price processes on the liquidation strategy.

The correlation of the two price processes is determined by the four volatility parameters $\sigma^{11}, \sigma^{12}, \sigma^{21}$, and $\sigma^{22}$. Fig. 3 shows the relationship between volatility components and the correlation of two asset prices.

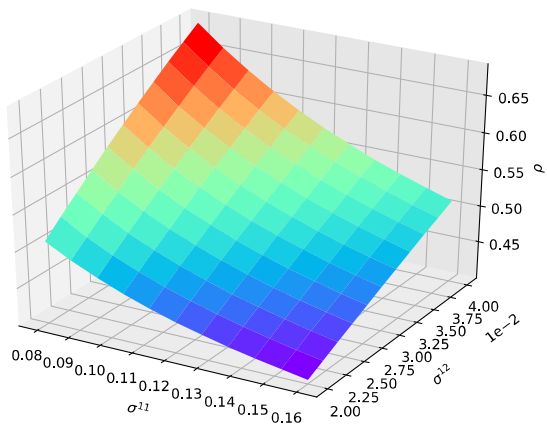 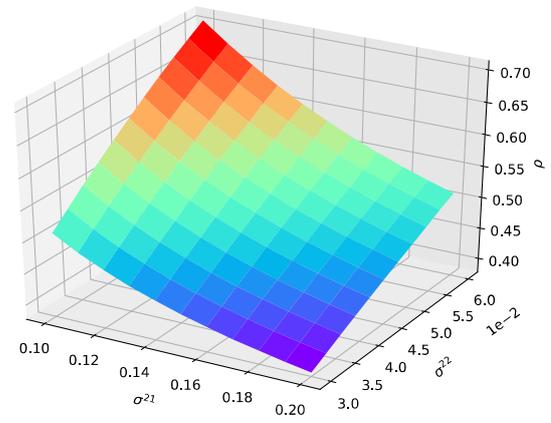

Fig. 3(a) Relationship between volatility components    Fig. 3(b) Relationship between volatility components
   of asset 1 and correlation coefficient of two assets      of asset 2 and correlation coefficient of two assets

Fig. 3 Relationship between volatility components of two assets and correlation coefficient of two assets

In Fig. 3, the x-axis and the y-axis respectively represent the volatility parameters, and the z-axes represent the correlation coefficient of two assets

It can be seen from Fig. 3 that the volatility components $\sigma^{11}, \sigma^{21}$ are negatively correlated with the correlation coefficient of the two assets, and $\sigma^{12}, \sigma^{22}$ are positively correlated with the correlation coefficient. When $\sigma^{11}, \sigma^{12}$ are 0.16, 0.02, the correlation coefficient between the two assets reaches the minimum value of 0.40, and when $\sigma^{11}, \sigma^{12}$ are 0.08 and 0.04, the correlation coefficient reaches the maximum value of 0.69. When $\sigma^{21}, \sigma^{22}$ are 0.2 and 0.03, the correlation coefficient reaches the minimum value of 0.38, and when $\sigma^{21}, \sigma^{22}$ are 0.1 and 0.06, the correlation coefficient reaches the maximum value of 0.71. Through subsequent analysis, it can be concluded that the absolute value of the difference between the respective volatility components of assets is negatively correlated with the correlation coefficient of the two assets, which will be displayed later.

To discuss the impact of the internal correlation of two assets on the optimal liquidation time and average total liquidation cost rate, we simulate and analyze the problem by fixing the volatility components of one asset and changing the other ones'.

To prevent the correlation coefficient of the two assets from being affected by the size of the volatility components of asset 2, we fix volatility components of asset 2 $\sigma^{21}, \sigma^{22}$ both at 0.01. In addition, in order to ensure that the optimal liquidation time and average total liquidation cost rate avoid being affected by the size of the volatility of asset 1, we fix the volatility of asset 1 $\sigma^1$ at 0.4, set the initial value of volatility component $\sigma^{11}$ to 0.04, and gradually increase it by 10% of the initial value and adjust it for 99 times, while the volatility component $\sigma^{12}$ is given by the following equation:

$$\sigma^{12} = \sqrt{(\sigma^1)^2 - (\sigma^{12})^2}$$

Other parameters remain unchanged, namely, the temporary impact coefficients $\eta^1, \eta^2$ are $3\times10^{-8}$ and $5\times10^{-8}$, the permanent impact coefficients of themselves $\gamma^{11}, \gamma^{22}$ are $3\times10^{-9}$ and $5\times10^{-9}$, and the external permanent impact coefficients $\gamma^{12}, \gamma^{21}$ are $1\times10^{-9}$ and $2\times10^{-9}$. Thus we can obtain 100 cases. In each case we run 1,000 simulations and count the results.

Table 3 gives the settings of volatility and volatility components in simulation.

Table 3 Settings of volatility and volatility components

|  | Initial Value | Adjustment Range | Adjustment Times | Final Value |
|---|---|---|---|---|
| $\sigma^1$ | 0.5 | 0 | 0 | 0.5 |
| $\sigma^{11}$ | 0.04 | 0.004 | 99 | 0.436 |
| $\sigma^{12}$ | 0.498 | - | 99 | 0.245 |
| $\sigma^{21}$ | 0.01 | - | 0 | 0.01 |
| $\sigma^{22}$ | 0.01 | - | 0 | 0.01 |

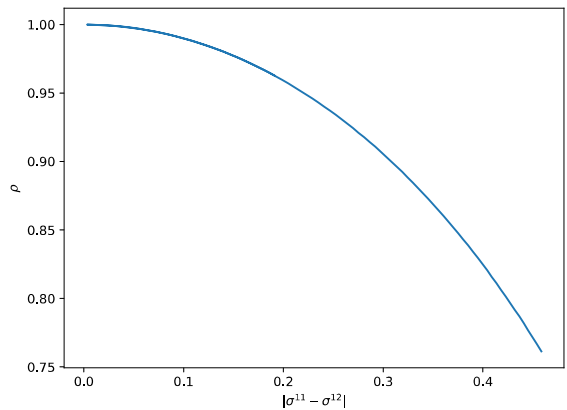 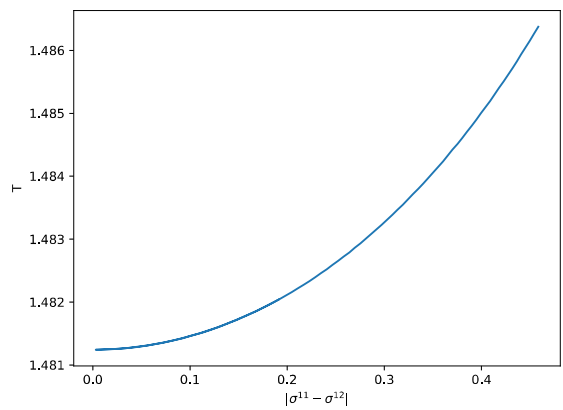

Fig. 4(a) Relationship between the absolute value of the difference in the volatility components of asset 1 and correlation coefficient of two assets

Fig. 4(b) Relationship between the absolute value of the difference in the volatility components of asset 1 and optimal liquidation time

Fig. 4 Relationship between the absolute value of the difference in the volatility components of asset 1 and correlation coefficient of two assets and optimal liquidation time

In Fig. 4, the x-axis represent the absolute values of the difference in the volatility components of asset 1, while the y-axis in Fig. 4(a) represents the correlation coefficient of two assets and the y-axis in Fig. 4(b) represents the optimal liquidation time.

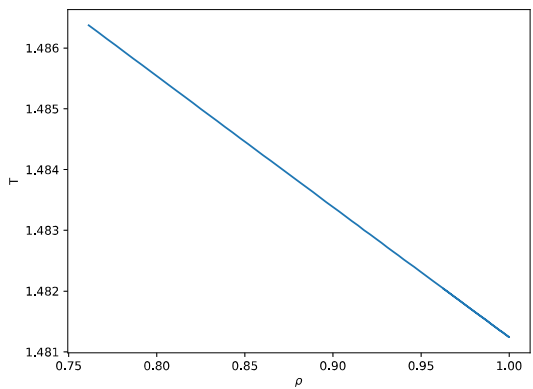 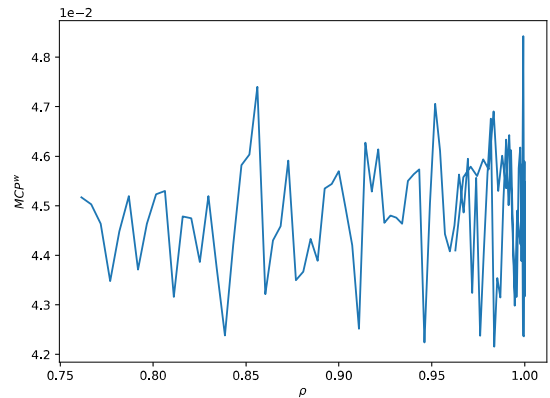

Fig. 5(a) Relationship between asset correlation coefficient and optimal liquidation time

Fig. 5(b) Relationship between asset correlation coefficient and average total liquidation cost rate

Fig. 5 Relationship between asset correlation coefficient and optimal liquidation time and average total liquidation cost rate

In Fig. 5, the x-axis represent the correlation coefficients of two assets, while the y-axis in Fig. 5(a) represents the optimal liquidation time and the y-axis in Fig. 5(b) represents the average total liquidation cost rate

From Fig. 4 and Fig. 5, we can see that the absolute value of the difference in $\sigma^{11}, \sigma^{12}$ is negatively correlated with the correlation coefficient of two assets and is positively correlated with the optimal liquidation time. The correlation coefficient of assets shows a significant negative correlation with the optimal liquidation time, but there is no significant relationship with average total liquidation cost rate. In addition, in this process, the optimal liquidation time changes less.

When the absolute value of the difference in $\sigma^{11}, \sigma^{12}$ reaches the minimum value of 0.0031, the correlation coefficient of the asset reaches the maximum value of 0.9999, and the optimal liquidation time reaches the minimum value of 1.4812. When the absolute value of the difference in $\sigma^{11}, \sigma^{12}$ reaches the maximum value of 0.4584, the correlation coefficient of the assets reaches the minimum value of 0.7614, and the optimal liquidation time reaches the maximum value of 1.4836.

### 3.3 The effect of the permanent impact coefficients on asset liquidation

According to the model, there is no direct relationship between the permanent impact coefficients and the optimal liquidation time. Hence, we only discuss the effect of the external impact coefficients in the asset process on the liquidation cost.

Table 4 Settings of external permanent impact coefficients

|  | **Initial Value** | **Adjustment Range** | **Adjustment Times** | **Final Value** |
|---|---|---|---|---|
| $\gamma^{12}$ | 1×10⁻⁹ | 1×10⁻¹⁰ | 10 | 2×10⁻⁹ |
| $\gamma^{21}$ | 2×10⁻⁹ | 2×10⁻¹⁰ | 10 | 4×10⁻⁹ |

The settings of permanent impact coefficients are shown in Table 4. The external permanent impact coefficients $\gamma^{12}, \gamma^{21}$ are 1×10⁻⁹ and 2×10⁻⁹, respectively. We gradually increase them by 10% of the initial values, and adjust them 10 times, thus obtaining 121 cases. Other parameters remain the same, namely, the temporary impact coefficients $\eta^1, \eta^2$ are 3×10⁻⁸ and 5×10⁻⁸, the volatility components of asset 1 $\sigma^{11}, \sigma^{12}$ are 0.08 and 0.02, the volatility components of asset 2 $\sigma^{21}, \sigma^{22}$ are 0.1 and 0.03, and the self-permanent impact coefficients $\gamma^{11}, \gamma^{22}$ are 3×10⁻⁹ and 5×10⁻⁹. Each case is simulated 1,000 times and the results are shown in Fig. 6.

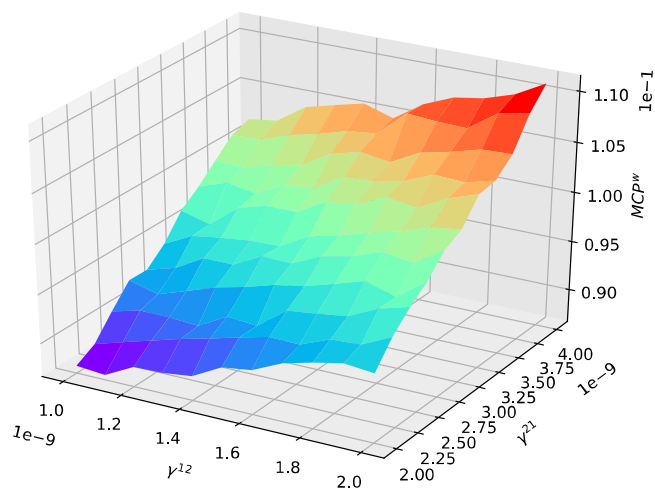

Fig. 6 Relationship between the external permanent impact coefficients of two assets and the average total liquidation cost rate

In Fig. 6, the x-axis and the y-axis respectively represent external permanent impact coefficients of two assets $\gamma^{12}, \gamma^{21}$, and the z-axis represents the average total liquidation cost rate

The results show that the average total liquidation cost rate increases with the increase of the external permanent impact coefficients of the two assets, showing a significant positive correlation.

4. **Conclusion**

This paper discusses a kind of portfolio liquidation problem under discrete time. The problem we ask is to liquidate the N risky assets (portfolio) for a given initial position. First, we assume that the asset price process is generated by an independent N-dimensional standard Brownian motion. Second, we assume that the N assets adopt a synchronous equal-interval and equal-share ordering strategy. Third, we assume that the permanent and temporary price impacts generated during the liquidation process are linear functions of transaction volume, the permanent impact will induce an impact on the prices of other assets, and the temporary impact only affects the price of the asset itself. Fourth, we set up the optimal liquidation strategy model under the VaR risk criterion and give the optimal liquidation time of the portfolio. Finally, we analyze the liquidation process of a portfolio composed of two assets through a simulation method and discuss the impact of different parameters on the liquidation strategy and liquidation cost.

The liquidation strategy of portfolio is a hot topic in the field of quantitative finance. This kind of problem usually requires solving the HJB equation under certain conditions, but generally can only result in a numerical solution. In contrast, the method presented in this paper has certain practicality. Nevertheless, the real asset prices show a significant external jump phenomenon, and some papers note that the form of price shock is usually nonlinear and elastic attenuation, therefore, this issue deserves further study, which is also the direction of our future work.